
\magnification=1200
\baselineskip=18pt

\def\[{\c c}
\def\\{\'\i }
\def\sqr#1#2{{\vcenter{\hrule height.#2pt
     \hbox{\vrule width.#2pt height#1pt \kern#1pt
      \vrule width.#2pt} \hrule height.#2pt}}}

\font\titlea=cmb10 scaled\magstep1

\font\titlec=cmb10 scaled\magstep3

\hfill IF-UFRJ-20/94
\bigskip
\centerline{\titlec BFT quantization of chiral-boson theories}
\vskip 2.0 cm
\centerline{R. Amorim$\,^\ast$ and J. Barcelos-Neto$\,^\dagger$}
\bigskip
\centerline{\it Instituto de F\\sica}
\centerline{\it Universidade Federal do Rio de Janeiro}
\centerline{\it RJ 21945-970 - Caixa Postal 68528 - Brasil}
\vskip 1.5 cm
\centerline{\titlea Abstract}
\bigskip
We use the method due to Batalin, Fradkin and Tyutin (BFT) for the
quantization of chiral boson theories. We consider the
Floreanini-Jackiw (FJ) formulation as well as others with linear
constraints.

\vfill
\bigskip
\hbox to 3.5 cm {\hrulefill}\par
\item{($\ast$)} Electronic mail: ift01001@ufrj
\item{($\dagger$)} Electronic mail: ift03001@ufrj and
barcelos@vms1.nce.ufrj.br
\eject

{\titlea I. Introduction}
\bigskip
Chiral-bosons are relevant for the understanding of several models
with intrinsical chirality. Among them we find superstrings,
W-gravities and general two-dimensional field theories in the
light-cone. Although apparently simple, chiral-bosons present
intriguing and interesting features. The great amount of work done on
this subject reveals its polemic character.

\medskip
One of the important points concerning chiral-bosons is related to
its chirality constraint. In the FJ model~[1], there is just one
continuous second-class constraint~[2], which contains however a zero
mode that causes a time dependent symmetry in its action~[3]. One
alternative way of introducing the chiral constraint in the
two-dimensional scalar field theory is by means of a Lagrange
multiplier~[4]. However, this model does not appear to be equivalent
to the FJ one~[5], even though both of them exhibit the same chiral
constraint as classical equation of motion. In a more recent work, it
has been shown how the chiral constraint can be correctly implemented
by means of Lagrange multipliers in order to be equivalent to the FJ
model~[6,7].

\medskip
All these facts make the constraint structure of chiral-boson
theories an interesting subject. In addition we could also mention
the Siegel formulation~[8], where the chiral-constraint appears as
first-class. On the other hand, the quantization methods of
constrained theories has been improved in these last two decades~[9].
One of these methods, due to Batalin, Fradkin and Tyutin
(BFT)~[10,11] has as main purpose the transformation of second-class
constraints into first-class ones. This is achieved with the aid of
auxiliary fields that extend the phase space in a convenient way.
After that, we have a gauge invariant system which matches the
original theory when the so-called unitary gauge is chosen.

\medskip
The BFT method is quite elegant and operates systematically. Our
purpose in this paper is to introduce it in chiral-boson theories. In
order to emphasize and clarify some particularities of the method, we
make a brief report of it in the Sec.~II. In Sec.~III we apply it to
the FJ chiral-boson. The main problem to be circumvented is that the
BFT method as originally introduced assumes that the system
contains an even number of second-class constraints. The FJ theory
has just one (continuous) constraint. So, one can take it as an
infinite mode expansion that naturally gives an even (although
infinite) number of second-class constraints. This way of implement
the BFT quantization method over the FJ chiral-boson has been
recently treated~[12]. Here we consider the BFT method by directly
using the continuous constraint.

\medskip
In Sec.~IV, we consider the chiral-boson formulated by means of
linear constraint~[4]. The problem mentioned above does not occur
here because there is a pair of second-class constraints. The
analysis of this model has also been recently reported in
literature~[13] where it was found two Wess-Zumino Lagrangians. We
have not found the same result here. The main difference of our
approaches is that we have considered the so-obtained first-class
constraints in a strongly involutive way, as stablishes the BFT
method, and not weakly as done in ref.~[13].  To conclude, we
consider in Sec.~V an alternative way of introducing the linear
constraint in the chiral boson theory~[6,7] and show that the final
quantum result is equivalent to the FJ one. We devote Sec.~VI to some
concluding remarks.

\vskip 1cm
{\titlea II. Brief review of the BFT formalism}
\bigskip
Let us consider a system described by a Hamiltonian $H_0$ in a
phase-space $(q^i,p_i)$ with $i=1,\dots,N$. Let us suppose that the
coordinates are bosonic (extension to include fermionic degrees of
freedom and to the continuous case can be done in a straightforward
way). Let us also suppose that there just exist second-class
constraints (at the end of this section we refer to the case where
first-class constraints are also present). Denoting them by $T_a$,
with $a=1,\dots M<2N$, we have

$$\bigl\{T_a,\,T_b\bigr\}=\Delta_{ab}\,,\eqno(2.1)$$

\bigskip\noindent
where $\det(\Delta_{ab})\not=0$.

\medskip
The general procedure of the BFT formalism is to convert second-class
constraints into first-class ones. This is achieved by introducing
auxiliary canonical variables, one for each second-class constraint
(the connection between the number of second-class constraints and
the new variables in one-to-one is to keep the same number of the
physical degrees of freedom in the resulting extended theory).
Denoting these auxiliary variables by $\psi^a$ we assume that they
have the following general structure

$$\bigl\{\psi^a,\,\psi^b\bigr\}=\omega^{ab}\,,\eqno(2.2)$$

\bigskip\noindent
where $\omega^{ab}$ is a constant quantity with
$\det\,(\omega^{ab})\not=0$. The obtainment of these quantities is
discussed in what follows. It is embodied in the obtainment of the
resulting first-class constraints that we denote by $\tilde T_a$. Of
course, these depend on the new variables $\psi^a$, namely

$$\tilde T_a=\tilde T_a(q,p;\psi)\eqno(2.3)$$

\bigskip\noindent
and satisfy the boundary condition

$$\tilde T_a(q,p;0)=\tilde T_a(q,p)\,.\eqno(2.4)$$

\bigskip\noindent
Another characteristic of these new constraints is that they are
assumed to be strongly involutive, i.e.

$$\bigl\{\tilde T_a,\,\tilde T_b\bigr\}=0\,.\eqno(2.5)$$

\bigskip
The solution of (2.5) can be achieved by considering $\tilde T_a$
expanded as

$$\tilde T_a=\sum_{n=0}^\infty T_a^{(n)}\,,\eqno(2.6)$$

\bigskip\noindent
where $T_a^{(n)}$ is a term of order $n$ in $\psi$. Compatibility
with the boundary condition (2.4) requires that

$$T_a^{(0)}=T_a\,.\eqno(2.7)$$

\bigskip\noindent
The replacement of (2.6) into (2.5) leads to a set of equations, one
for each coefficient of $\psi^n$. We list below some of them

$$\eqalignno{\bigl\{T_a^{(0)},\,T_b^{(0)}\bigr\}_{(q,p)}
&+\bigl\{T_a^{(1)},\,T_b^{(1)}\bigr\}_{(\psi)}=0\,,&(2.8a)\cr
\bigl\{T_a^{(0)},\,T_b^{(1)}\bigr\}_{(q,p)}
&+\bigl\{T_a^{(1)},\,T_b^{(0)}\bigr\}_{(q,p)}
+\bigl\{T_a^{(1)},\,T_b^{(2)}\bigr\}_{(\psi)}
+\bigl\{T_a^{(2)},\,T_b^{(1)}\bigr\}_{(\psi)}=0\,,&(2.8b)\cr
\bigl\{T_a^{(0)},\,T_b^{(2)}\bigr\}_{(q,p)}
&+\bigl\{T_a^{(1)},\,T_b^{(1)}\bigr\}_{(q,p)}
+\bigl\{T_a^{(2)},\,T_b^{(0)}\bigr\}_{(q,p)}
+\bigl\{T_a^{(1)},\,T_b^{(3)}\bigr\}_{(\psi)}\cr
&+\bigl\{T_a^{(2)},\,T_b^{(2)}\bigr\}_{(\psi)}
+\bigl\{T_a^{(3)},\,T_b^{(1)}\bigr\}_{(\psi)}=0\,,&(2.8c)\cr
&\vdots\cr}$$

\bigskip\noindent
These correspond to coefficients of the powers $\psi^0$, $\psi^1$,
$\psi^2$, $\dots$, respectively. The notation used above,
$\{,\}_{(q,p)}$ and $\{,\}_{(\psi)}$, represent the parts of the
Poisson bracket $\{,\}$ relative to the variables $(q,p)$ and
$(\psi)$.

\medskip
Equations (2.8) are used iteratively in the obtainment of the
corrections $T^{(n)}$ ($n\geq1$). The first equation (2.8) shall give
$T^{(1)}$. With this result and (2.8{\it b}), one calculates
$T^{(2)}$, and so on. Since $T^{(1)}$ is linear in $\psi$ we may
write

$$T_a^{(1)}=X_{ab}(q,p)\,\psi^b\,.\eqno(2.9)$$

\bigskip\noindent
Introducing this expression into (2.8{\it a}) and using the boundary
condition (2.4), as well as (2.1) and (2.2), we get

$$\Delta_{ab}+X_{ac}\,\omega^{cd}\,X_{bd}=0\,.\eqno(2.10)$$

\bigskip\noindent
We notice that this equation does not give $X_{ab}$ univocally,
because it also contains the still unknown $\omega^{ab}$. What we
usually do is to choose $\omega^{ab}$ in such a way that the new
variables are
\eject\noindent
unconstrained
\footnote{(*)}{It is opportune to mention that this procedure is not
always possible to be done. We shall return to this point in the
examples to be discussed in the next sections.}.
The knowledge of $X_{ab}$ permits us to obtain $T_a^{(1)}$. If
$X_{ab}$ do not depend on $(q,p)$, it is easily seen that
$T_a+T_a^{(1)}$ is already strongly involutive. When this occurs, we
are succeed in obtaining $\tilde T_a$. If this is not so we have to
introduce $T_a^{(1)}$ into (2.8{\it b}) to calculate $T_a^{(2)}$,
and so on.

\medskip
Another point in the BFT formalism is that any dynamical function
$A(q,p)$ (for instance the Hamiltonian) has also to be properly
modified in order to be strongly involutive with the first-class
constraints $\tilde T_a$. Denoting the modified quantity by $\tilde
A(q,p;\psi)$, we then have

$$\bigl\{\tilde T_a,\,\tilde A\bigr\}=0\,.\eqno(2.11)$$

\bigskip\noindent
In addition, $\tilde A$ has also to satisfy  the boundary condition

$$\tilde A(q,p;0)=A(q,p)\,.\eqno(2.12)$$

\bigskip
The obtainment of $\tilde A$ is similar to what was done to get
$\tilde T_a$, that is to say, we consider an expansion like

$$\tilde A=\sum_{n=0}^\infty A^{(n)}\,,\eqno(2.13)$$

\bigskip\noindent
where $A^{(n)}$ is also a term of order $n$ in $\psi's$.
Consequently, compatibility with (2.12) requires that

$$A^{(0)}=A\,.\eqno(2.14)$$

\vfill\eject
\noindent The combination of (2.6), (2.11) and (2.13) gives

$$\eqalignno{\bigl\{T_a^{(0)},\,A^{(0)}\bigr\}_{(q,p)}
&+\bigl\{T_a^{(1)},\,A^{(1)}\bigr\}_{(\psi)}=0\,,&(2.15a)\cr
\bigl\{T_a^{(0)},\,A^{(1)}\bigr\}_{(q,p)}
&+\bigl\{T_a^{(1)},\,A^{(0)}\bigr\}_{(q,p)}
+\bigl\{T_a^{(1)},\,A^{(2)}\bigr\}_{(\psi)}
+\bigl\{T_a^{(2)},\,A^{(1)}\bigr\}_{(\psi)}=0\,,&(2.15b)\cr
\bigl\{T_a^{(0)},\,A^{(2)}\bigr\}_{(q,p)}
&+\bigl\{T_a^{(1)},\,A^{(1)}\bigr\}_{(q,p)}
+\bigl\{T_a^{(2)},\,A^{(0)}\bigr\}_{(q,p)}
+\bigl\{T_a^{(1)},\,A^{(3)}\bigr\}_{(\psi)}\cr
&+\bigl\{T_a^{(2)},\,A^{(2)}\bigr\}_{(\psi)}
+\bigl\{T_a^{(3)},\,A^{(1)}\bigr\}_{(\psi)}=0\,,&(2.15c)\cr
&\vdots\cr}$$

\bigskip\noindent
which correspond to the coefficients of the powers $\psi^0$,
$\psi^1$, $\psi^2$, etc., respectively. The first expression above
gives us $A^{(1)}$

$$A^{(1)}=-\,\psi^a\,\omega_{ab}\,X^{bc}\,
\bigl\{T_c,\,A\bigr\}\,,\eqno(2.16)$$

\bigskip\noindent
where $\omega_{ab}$ and $X^{ab}$ are the inverse of $\omega^{ab}$ and
$X_{ab}$.

\medskip
In the obtainment of $T^{(1)}_a$ we had seen that $T_a+T^{(1)}_a$ was
strongly involutive if the coefficients $X_{ab}$ do not depend on
$(q,p)$. However, the same argument does not necessarily apply here.
It might be necessary to calculate other corrections to obtain the
final $\tilde A$. Let us discuss how this can be systematically done.
We consider the general case first.  The correction $A^{(2)}$ comes
from (2.15{\it b}), that we conveniently rewrite as

$$\bigl\{T_a^{(1)},\,A^{(2)}\bigr\}_{(\psi)}
=-\,G_a^{(1)}\,,\eqno(2.17)$$

\bigskip\noindent
where

$$G_a^{(1)}=\bigl\{T_a,\,A^{(1)}\bigr\}_{(q,p)}
+\bigl\{T_a^{(1)},\,A\bigr\}_{(q,p)}
+\bigl\{T_a^{(2)},\,A^{(1)}\bigr\}_{(\psi)}\,.\eqno(2.18)$$

\bigskip\noindent
Thus

$$A^{(2)}=-{1\over2}\,\psi^a\,\omega_{ab}\,X^{bc}\,G_c^{(1)}\,.
\eqno(2.19)$$

\bigskip\noindent
In the same way, other terms can be obtained. The final general
expression reads

$$A^{(n+1)}=-{1\over n+1}\,\psi^a\,\omega_{ab}\,X^{bc}\,G_c^{(n)}\,,
\eqno(2.20)$$

\bigskip\noindent
where

$$G_a^{(n)}=\sum_{m=0}^n\bigl\{T_a^{(n-m)},\,A^{(m)}\bigr\}_{(q,p)}
+\sum_{m=0}^{n-2}\bigl\{T_a^{(n-m)},\,A^{(m+2)}\bigr\}_{(\psi)}
+\bigl\{T_a^{(n+1)},\,A^{(1)}\bigr\}_{(\psi)}\,.\eqno(2.21)$$

\bigskip\noindent
For the particular case when $X_{ab}$ do not depend on $(q,p)$ we
have that the corrections $A^{(n+1)}$ can be obtained by the same
expression (2.20), but $G_a^{(n)}$ simplifies to

$$G_a^{(n)}=\bigl\{T_a,\,A^{(n)}\bigr\}_{(q,p)}
+\bigl\{T_a^{(1)},\,A^{(n-1)}\bigr\}_{(q,p)}\,.\eqno(2.22)$$

\bigskip
To conclude this brief report on the BFT formalism, we refer to the
case where there are also first-class constrains. Let us call them by
$F_\alpha$. We consider that the constraints of the theory satisfy
the following involutive algebra (with the use of the Dirac bracket
definition to strongly eliminate the second-class contraints)

$$\eqalignno{\bigl\{F_\alpha,\,F_\beta\bigr\}_D
&=U_{\alpha\beta}^\gamma\,F_\gamma
+I_{\alpha\beta}^a\,T_a\,,&(2.23a)\cr
\bigl\{H_0,\,F_\alpha\bigr\}_D
&=V_\alpha^\beta\,F_\beta+K_\alpha^a\,T_a\,,&(2.23b)\cr
\bigl\{F_\alpha,\,T_b\bigr\}_D
&=0\,.&(2.23c)\cr}$$

\bigskip\noindent
In (2.23), $U_{\alpha\beta}^\gamma$, $I_{\alpha\beta}^a$,
$V_\alpha^\beta$ and $K_\alpha^a$ are structure functions of the
involutive algebra.

\medskip
The BFT procedure in this case also introduces one auxiliary variable
for each one of the second-class constraints (this is also in
agreement with the counting of the physical degrees of freedom of the
initial theory). All the constraints and the Hamiltonian have to be
properly modified in order to satisfy the same involutive algebra
above, namely,

$$\eqalignno{\bigl\{\tilde F_\alpha,\,\tilde F_\beta\bigr\}_D
&=U_{\alpha\beta}^\gamma\,\tilde F_\gamma
+I_{\alpha\beta}^a\,\tilde T_a\,,&(2.24a)\cr
\bigl\{\tilde H_0,\,\tilde F_\alpha\bigr\}_D
&=V_\alpha^\beta\,\tilde F_\beta
+K_\alpha^a\,\tilde T_a\,,&(2.24b)\cr
\bigl\{\tilde F_\alpha,\,\tilde T_b\bigr\}_D
&=0\,.&(2.24c)\cr}$$

\bigskip\noindent
Since the algebra is now weakly involutive, the iterative calculation
of the previous case cannot be applied here. We have to figure out
the corrections that have to be done in the initial quantities.

\vskip 1cm
{\titlea III. The Floreanini-Jackiw chiral boson}
\bigskip
This is described by the following Lagrangian density~[1]

$${\cal L}=\dot\phi\,\phi^\prime-\phi^{\prime2}\,,\eqno(3.1)$$

\bigskip\noindent
where dot and prime represent derivatives with respect time and space
coordinates respectively. Space-time is assumed to be a
two-dimensional Minkowskian variety. The chiral condition
$\dot\phi-\phi^\prime=0$ is obtained as an equation of motion up to a
general function of time. The canonical momentum conjugate to $\phi$
is

$$\pi=\phi^\prime\,.\eqno(3.1)$$

\bigskip\noindent
This is a constraint that we denote by

$$T(\phi,\pi)=\pi-\phi^\prime\,.\eqno(3.2)$$

\bigskip\noindent
We construct the primary Hamiltonian density

$$\eqalignno{{\cal H}&=\pi\dot\phi-{\cal L}+\xi T\,,\cr
&=\bigl(\pi-\phi^\prime\bigr)\dot\phi+\phi^{\prime2}
+\xi\bigl(\pi-\phi^\prime\bigr)\,,\cr
&\rightarrow\phi^{\prime2}
+\xi\bigl(\pi-\phi^\prime\bigr)\,,&(3.3)\cr}$$

\bigskip\noindent
where in the last step we have absorbed the velocity $\dot\phi$ in
the Lagrange multiplier $\xi$. The consistency condition for the
constraint $T$ does not lead to any new one.

\medskip
The constraint above is second-class, in a sense that they satisfy
the Poisson bracket relation

$$\bigl\{T(x),\,T(y)\bigr\}=-\,2\,\delta^\prime(x-y)\,.\eqno(3.4)$$

\bigskip\noindent
This bracket and those that follow are taken at the same time,
$x_0=y_0$.

\medskip
To implement the BFT formalism we have to introduce an auxiliary
field $\psi$ satisfying the following general bracket structure

$$\bigl\{\psi(x),\,\psi(y)\bigr\}=\omega(x,y)\,,\eqno(3.5)$$

\bigskip\noindent
where $\omega$ is antisymmetric in $x$, $y$. Of course, this cannot
be achieved in terms of Poisson brackets.  There are two ways (maybe
more) to circumvent this problem. One of them is considering all
fields expanded in terms of Fourier modes.  This leads to a infinite
number of variables and it is possible to have an expression like
(3.5) but in terms of Poisson brackets. As it was told in the
introduction, we have discussed this case in a previous paper~[12].
The second procedure is to consider that $\psi$ is constrained and
that expression (3.5) is realized in terms of Dirac brackets. We
shall follow this second possibility here.

\medskip
We now extend  the constraint $T$ to $\tilde T$

$$\tilde T=\tilde T(\phi,\pi;\psi)\,,\eqno(3.6)$$

\bigskip\noindent
with the boundary condition

$$\tilde T(\phi,\pi;0)=T\eqno(3.7)$$

\bigskip\noindent
and consider that they are strong involutive, i.e.

$$\bigl\{\tilde T(x),\,\tilde T(y)\bigr\}=0\,.\eqno(3.8)$$

\bigskip
The obtainment of $\tilde T$ follows the procedure discussed  in
Sec.~2. We first have to solve the equation (see expression 2.10)

$$\int dzdr\,X(x,z)\,w(z,r)\,X(y,r)=-\Delta(x,y)\,,\eqno(3.9)$$

\bigskip\noindent
where $\Delta(x,y)$ is related to the structure given by (3.4), i.e.

$$\Delta(x,y)=-\,2\,\delta^\prime(x-y)\,.\eqno(3.10)$$

\bigskip\noindent
Since we are considering that the bracket involving $\psi$'s is
constrained, let us choose

$$\omega(x,y)=2\,\delta^\prime(x-y)\,.\eqno(3.11)$$

\bigskip\noindent
In consequence, the solution of (3.9) gives

$$X(x,y)=\delta(x-y)\,.\eqno(3.12)$$

\bigskip\noindent
We notice that the quantity $X$ does not depend on the initial fields
$(\phi,\pi)$. This means that

$$\eqalignno{\tilde T(x)&=T(x)+T^{(1)}(x)\,,\cr
&=T(x)+\int dy\,X(x,y)\,\psi(y)\,,\cr
&=T(x)+\psi(x)\,.&(3.13)\cr}$$

\bigskip\noindent
In fact, one can easily verify that $\tilde T=T+\psi$ satisfies the
involutive expression (3.8).

\medskip
We now pass to consider the obtainment of $\tilde H_c$. Considering
what we have seen in Sec.~2 and that we also have $T^{(n)}=0$ for
$n\geq2$, the corrections that give $\tilde H_c$ can be written as

$$H_c^{(n+1)}=-{1\over n+1}\,
\int dxdydz\,\,\psi(x)\,\omega^{-1}(x,y)\,
X^{-1}(y,z)\,G^{(n)}(z)\,,\eqno(3.14)$$

\bigskip\noindent
where $G^{(n)}(x)$ is given by

$$G^{(n)}(x)=\bigl\{T(x),\,H_c^{(n)}\bigr\}_{(\phi,\pi)}
+\bigl\{T^{(1)}(x),\,H_c^{(n-1)}\bigr\}_{(\phi,\pi)}
\eqno(3.15)$$

\bigskip\noindent
and

$$\eqalignno{\omega^{-1}(x,y)&={1\over2}\,\theta(x-y)\,,\cr
X^{-1}(x,y)&=\delta(x-y)\,.&(3.16)\cr}$$

\bigskip\noindent
The quantity $\theta(x-y)$ that appears in (3.16) is the usual
theta-function. The initial canonical Hamiltonian can be figure out
from (3.3) as

$$H_c=\int dx\,\phi^{\prime2}\,.\eqno(3.17)$$

\bigskip\noindent
Thus, using (3.15) and (3.17) we get

$$G^{(0)}(x)=2\,\phi^{\prime\prime}\,.\eqno(3.18)$$

\bigskip\noindent
The combination of (3.14), (3.17) and (3.18) permit us to calculate
$H_c^{(1)}$

$$H_c^{(1)}=-\int dx\,\psi\,\phi^\prime\,.\eqno(3.19)$$

\bigskip\noindent
The next corrections are

$$\eqalignno{G^{(1)}&=\psi^\prime(x)\,,&(3.20)\cr
H_c^{(2)}&={1\over4}\,\int dx\,\psi^2(x)\,.&(3.21)\cr}$$

\bigskip\noindent
It is easily seen that other corrections are zero. The extended
canonical Hamiltonian density is then given by

$$\tilde{\cal H}_c=\phi^{\prime2}-\psi\,\phi^\prime
+{1\over4}\,\psi^2\,.\eqno(3.22)$$

\bigskip\noindent
In fact, one can check that it has strong involution with the
constraint $\tilde T$.

\medskip
We now look for the Lagrangian that leads to this extended theory. A
consistent way of doing this is by means of the path integral
formalism in the Faddeev-Senjanovic procedure~[14,15]. Since $\psi$
is constrained, it is not difficult to see that the constraint that
leads to the bracket (3.5) with $\omega$ given by (3.11) is

$$R=p_\psi^\prime+{1\over4}\,\psi\,,\eqno(3.23)$$

\bigskip\noindent
where $p_\psi$ is the canonical momentum related to $\psi$. The
general expression of the vacuum functional then reads

$$Z=N\int[d\mu]\exp\,\Bigl\{i\int d^2x\,
\bigl(p\dot\lambda+\pi\dot\phi+p_\psi\dot\psi
-\tilde{\cal H}_c\bigr)\Bigr\}\,,\eqno(3.24)$$

\bigskip\noindent
with the measure $[d\mu]$ given by

$$[d\mu]=[d\phi][d\psi][d\pi][dp_\psi]\,
\delta[\pi-\psi^\prime+\psi]\,\delta[p_\psi^\prime+{1\over4}\psi]\,
\delta[\tilde\chi]\,\vert\det\{,\}\vert^{1/2}\eqno(3.25)$$

\bigskip\noindent
and $\tilde\chi$ is the gauge-fixing function corresponding to the
first-class constraint $\tilde T$. The term $\vert\det\{,\}\vert$ is
representing the determinant of all constraints of the theory,
including the gauge-fixing ones. The integration over $\pi$ is easily
done by using the first delta functional and the integration over
$p_\psi$ gives a nonlocal term. The final result is

$$\eqalignno{Z=N\int[d\phi][d\psi]\,\delta[\chi]\,
\vert\det\{,\}\vert^{1/2}\,
&\exp\,\Bigl\{i\int d^2x\,\Bigl[\dot\phi\phi^\prime
-\phi^{\prime2}+\psi\,\bigl(\phi^\prime-\dot\phi\bigr)
-{1\over4}\,\psi^2\cr
&-{1\over4}\,\dot\psi\int dy\,\theta(x-y)\,\psi(y)
\Bigr]\Bigr\}\,.&(3.26)\cr}$$

\bigskip\noindent
{}From the expression above we obtain that the Lagrangian density that
leads to the extended theory we have discussed is

$${\cal L}=\dot\phi\phi^\prime
-\phi^{\prime2}+\psi\,\bigl(\phi^\prime-\dot\phi\bigr)
-{1\over4}\,\psi^2-{1\over4}\,\dot\psi
\int dy\,\theta(x-y)\,\psi(y)\,.\eqno(3.27)$$

\bigskip\noindent
As one observes, it leads to the usual FJ Lagrangian of the
chiral-boson theory when the auxiliary field $\psi$ is turned off.
Incidentally we mention that this is the same Lagrangian obtained
from Fourier modes expansion as discussed in reference~[12].

\medskip
It may be borrowing us the presence of a second-class constraint in
the implementation of the BFT formalism. This constraint could also
be transformed into a first-class one by introducing a new auxiliary
field, also constrained. This last constraint can also be transformed
into first-class in a endless process~[16].

\vskip 1cm
{\titlea IV. Chiral-bosons with linear constraint}
\bigskip
The chiral-boson condition $\dot\phi-\phi^\prime=0$ can also be
introduced in the two-dimensional scalar field theory by means of
a Lagrange multiplier~[4]
\footnote{(*)}{We use the metric convention:
$\eta_{00}=-\eta_{11}=-1$, $\eta_{01}=0$.}

$${\cal L}=-{1\over2}\,\partial_\mu\phi\partial^\mu\phi
+\lambda\bigl(\dot\phi-\phi^\prime\bigr)\,.\eqno(4.1)$$

\eject\noindent
This theory has been criticized by some authors~[5]. The main
arguments are that it does not lead to a positive definite
Hamiltonian and that its physical spectrum is just the vacuum state.
In fact, the theory described by the Lagrangian (4.1) is not
equivalent to the FJ one, even though both of them contain the same
chiral condition $\dot\phi-\phi^\prime=0$ as classical equation of
motion. We are going to study this point with details by means of the
BFT formalism.  In the next section we shall discuss what is missing
in the Lagrangian (4.1) to correct describe the usual chiral-boson
theory.

\medskip
{}From the Lagrangian (4.1) we get the momenta

$$\eqalignno{p&={\partial{\cal L}\over\partial\dot\lambda}=0\,,
&(4.2)\cr
\pi&={\partial{\cal L}\over\partial\dot\phi}=\dot\phi+\lambda\,.
&(4.3)\cr}$$

\bigskip\noindent
Expression (4.2) is a (primary) constraint. We then construct the
primary Hamiltonian

$${\cal H}={1\over2}\,\bigl(\pi^2+\phi^{\prime2}+\lambda^2\bigr)
+\lambda\bigl(\phi^\prime-\pi\bigr)+\xi p\,,\eqno(4.4)$$

\bigskip\noindent
where the velocity $\dot\lambda$ was absorbed by $\xi$. The
consistency condition leads to a new constraint

$$\pi-\phi^\prime-\lambda=0\,.\eqno(4.5)$$

\bigskip\noindent
Let us denote these constraints by

$$\eqalignno{T_1&=p\,,&(4.6)\cr
T_2&=\pi-\phi^\prime-\lambda\,.&(4.7)\cr}$$

\bigskip\noindent
They are second-class and their nonvanishing brackets read

$$\eqalignno{\bigl\{T_1(x),\,T_2(y)\bigr\}&=\delta(x-y)\,,\cr
\bigl\{T_2(x),\,T_2(y)\bigr\}&=-2\,\delta^\prime(x-y)\,.&(4.8)\cr}$$

\bigskip\noindent
{}From these results, we can see that, in fact, Lagrangian (4.1) cannot
lead to the same theory of the FJ one. The former has two degrees of
freedom given by the fields $\phi$ and its momentum $\pi$ and one
second-class constraint. So, it has just one physical degree of
freedom. The case with linear constraint has four degrees of freedom,
related to $\phi$, $\pi$, $\lambda$ and $p$ and two second class
constraints. So, differently from the first case, it has two physical
degrees of freedom.

\medskip
To implement the BFT formalism, we introduce two auxiliary fields
$\psi^a$, with $a$=1,2, satisfying the following structure

$$\bigl\{\psi^a(x),\,\psi^b(y)\bigr\}=\omega^{ab}(x,y)\,,\eqno(4.9)$$

\bigskip\noindent
where $\omega^{ab}(x,y)$ is also antisymmetric in $a$, $b$.
We thus extend the constraints $T_a$ to $\tilde T_a$ such that

$$\tilde T_a=\tilde T_a(\phi,\pi,\lambda,p;\psi^a)\,,\eqno(4.10)$$

\bigskip\noindent
with the usual boundary condition

$$\tilde T_a(\phi,\pi,\lambda,p;0)=T_a\eqno(4.11)$$

\bigskip\noindent
and consider that they are strong involutive

$$\bigl\{\tilde T_a,\,\tilde T_b\bigr\}=0\,.\eqno(4.12)$$

\bigskip
The obtainment of $\tilde T_a$ is given as discussed in Sec. 2 and
in the previous section. First, we have to solve the
equation

$$\int dzdr\,X_{ac}(x,z)\,\omega^{cd}(z,r)\,X_{bd}(y,r)
=-\,\Delta_{ab}(x,y)\,,\eqno(4.13)$$

\bigskip\noindent
where $\Delta_{ab}$ is related to the structure of the Poisson
brackets of the constraints $T_a$. From (4.8), we get
\footnote{(*)}{$\epsilon_{12}=-\epsilon^{12}=1$.}

$$\Delta_{ab}=\bigl(\epsilon_{ab}
-2\delta_{a2}\,\delta_{b2}\,\partial_x\bigr)\,
\delta(x-y)\,.\eqno(4.14)$$

\bigskip
Here, it is possible to consider that the auxiliary variables
$\psi^a$ are not constrained. We thus take

$$\omega^{ab}(x,y)=\epsilon^{ab}\,\delta(x-y)\,,\eqno(4.15)$$

\bigskip\noindent
that is to say, we are considering that one of the $\psi^a$ is the canonical
momentum of the other. The bracket (4.9), where $\omega^{ab}$ is
given by (4.15), is in a unconstrained symplectic form.
Introducing (4.14) and (4.15) into (4.13) we get

$$X_{ab}(x,y)=\bigl(\epsilon_{ab}
-\delta_{a2}\,\delta_{b2}\,\partial_x\bigr)\,
\delta(x-y)\,.\eqno(4.16)$$

\bigskip\noindent
We notice that $X_{ab}$ do not depend on the initial fields $\phi$,
$\lambda$, $\pi$ and $p$. This means that the extended first-class
constraints $\tilde T_a$ are just given by

$$\eqalignno{\tilde T_a(x)&=T_a(x)+T_a^{(1)}(x)\,,\cr
&=T_a(x)+\int dy\,X_{ab}(x,y)\,\psi^b(y)\,.&(4.17)\cr}$$

\eject\noindent
Hence,

$$\eqalignno{\tilde T_1&=p+\psi^2\,,&(4.18)\cr
\tilde T_2&=\pi-\phi^\prime-\lambda-\psi^1-\psi^{2\prime}\,.
&(4.19)\cr}$$

\bigskip\noindent
One can actually see that $\tilde T_1$ and $\tilde T_2$ satisfy the
strong involution relation (4.12).

\medskip
Considering what was done in the previous section, the corrections
that give $\tilde H_{\rm C}$ are

$$H_{\rm C}^{(n+1)}=-{1\over n+1}\,\int dxdydz\,
\psi^a(x)\,\omega_{ab}(x,y)\,X^{bc}(y,z)\,
G^{(n)}_c(z)\,,\eqno(4.20)$$

\bigskip\noindent
where $G_a^{(n)}(x)$ are

$$G_a^{(n)}(x)=\bigl\{T_a(x),\,H_{\rm C}^{(n)}\bigr\}_
{(\phi,\lambda,\pi,p)}+
\bigl\{T_a^{(1)}(x),\,H_{\rm C}^{(n-1)}\bigr\}_
{(\phi,\lambda,\pi,p)}\eqno(4.21)$$

\bigskip\noindent
and the inverses $\omega_{ab}$ and $X^{ab}$ read

$$\eqalignno{\omega_{ab}(x,y)
&=\epsilon_{ab}\,\delta(x-y)\,,&(4.22)\cr
X^{ab}(x,y)&=\bigl(\epsilon^{ab}
-\delta^{a1}\,\delta^{b1}\,\partial_x\bigr)\,
\delta(x-y)\,.&(4.23)\cr}$$

\bigskip\noindent
As the canonical Hamiltonian is (see expression
4.4)

$${\cal H}_{\rm C}={1\over2}\,\bigl(\pi^2+\phi^{\prime2}
+\lambda^2\bigr)+\lambda\,\bigl(\phi^\prime-\pi\bigr)\,,
\eqno(4.24)$$

\bigskip\noindent
we obtain

$$\eqalignno{G_1^{(0)}&=T_2\,,&(4.25)\cr
G_2^{(0)}&=-T_2^\prime+\lambda^\prime\,,&(4.26)\cr
{\cal H}_{\rm C}^{(1)}&=\psi^1\,T_2+\psi^2\,\lambda^\prime\,,&(4.27)\cr
G_1^{(1)}&=-\psi^1-\psi^{2\prime}\,,&(4.28)\cr
G_2^{(1)}&=2\psi^{1\prime}(x)\,,&(4.29)\cr
{\cal H}_{\rm C}^{(2)}&={1\over2}\bigl(\psi^1\psi^1
+2\psi^1\psi^{2\prime}+\psi^2\psi^{2\prime\prime}\bigr)\,.
&(4.30)\cr}$$

\bigskip\noindent
Other corrections are zero. We can thus write the extended canonical
Hamiltonian

$$\eqalignno{\tilde H_{\rm C}&=H_{\rm C}
+H_{\rm C}^{(1)}+H_{\rm C}^{(2)}\,,\cr
&=\int dx\,\Bigl[{1\over2}\,\bigl(\pi^2+\phi^{\prime2}
+\lambda^2\bigr)+\lambda\bigl(\phi^\prime-\pi\bigr)
-\psi^1\bigl(\pi-\phi^\prime-\lambda\bigr)
-\psi^2\lambda^\prime\cr
&\phantom{=\int dx\,\Bigl[{1\over2}\,\,\bigl(\pi^2+\phi^{\prime2}
+\lambda^2\bigr)}
+{1\over2}\,\bigl(\psi^1\psi^1+2\psi^1\psi^{2\prime}
+\psi^2\psi^{2\prime\prime}\bigr)\Bigr]\,.&(4.31)\cr}$$

\bigskip\noindent
It is also a direct check to verify that $\tilde H_{\rm C}$ is, in
fact, also strongly involutive with the constraints $\tilde T_a$.

\medskip
In a previous analysis of the chiral-boson theory with linear
constraint, Miao, Zhou and Liu~[13] have found two WZ Hamiltonians,
contrarily to what we have found here. The difference between our
results is mainly because they have started with weakly involutive
relations instead of strong ones.

\medskip
Finally, we look for the Lagrangian that leads to this extended
theory. We are now considering that $\psi^a$ are not constrained.
{}From the simplectic form $\{\psi^a,\,\psi^b\}=\epsilon^{ab}$ we can
write them as the canonical pair

$$\eqalignno{\psi^1&=p_\psi\,,\cr
\psi^2&=\psi\,.&(4.32)\cr}$$

\bigskip\noindent
Here, as there are just first-class constraints, the general
expression of the vacuum functional follows from the Faddeev path
integral formulation~[14], i.e.

$$Z=N\int[d\mu]\,\exp\,\Bigl\{i\int d^2x\,
\bigl(p\dot\lambda+\pi\dot\phi
+p_\psi\dot\psi-\tilde{\cal H}\bigr)\Bigr\}\,,\eqno(4.33)$$

\bigskip\noindent
where the measure $[d\mu]$ reads

$$[d\mu]=[d\phi][d\pi][d\lambda][dp][d\psi][dp_\psi]\,
\vert\det\bigl\{\tilde T_a,\,\tilde\chi_a\bigr\}\vert\,
\prod_{a=1}^2\delta[\tilde T_a]\delta[\chi_a]\,.
\eqno(4.34)$$

\bigskip\noindent
The quantities $\tilde\chi_a$ are gauge-fixing constraints (in the
Faddeev procedure, they have to satisfy
$\{\tilde\chi_a,\,\tilde\chi_b\}=0$).

\medskip
The effective Lagrangian is then obtained by integrating over the
momenta. The use of the delta functionals makes the integration over
$p$ and $\pi$ quite trivial. Integration over $p_\psi$ gives another
delta functional (in this last step we assume that the gauge-fixing
constraints do not contain $p_\psi$). The result is

$$\eqalignno{Z=N\int[d\phi][d\lambda][d\psi]\,
\delta[\dot\phi-\phi^\prime+\dot\psi&-\psi^\prime]\,
\exp\,\Bigl\{i\int d^2x\,
\Bigl[\psi\bigl(\lambda^\prime-\dot\lambda\bigr)\cr
&+\bigl(\psi^\prime-\dot\psi\bigr)
\bigl(\phi^\prime+\lambda+\psi^\prime\bigr)\Bigr]\Bigr\}\,,
&(4.35)\cr}$$

\bigskip\noindent
where it was used the expression of the delta functional in the
effective action.

\medskip
One observes once more that the chiral boson theory with linear
constraint as given by (4.1) is not effectively equivalent to the FJ
one. For instance, if one turns out the auxiliary field one obtains
an identically null Lagrangian. This kind of problem had already been
pointed out in a previous paper~[6].
\medskip
An interesting point is that even though the effective Lagrangian in
(4.35) leads to an identically null theory when the auxiliary field
is removed, we can verify that it also leads to the model we have
here discussed, that is to say, with the first-class constraints
$\tilde T_a$.

\vskip 1cm
{\titlea V. Chiral-bosons with linear constraint revisited}
\bigskip
There is another way of introducing the linear constraint in the
chiral-boson theory. Instead of the Lagrangian (4.1) we take~[6,7]

$${\cal L}=-{1\over2}\,\partial_\mu\phi\partial^\mu\phi
+\lambda\bigl(\dot\phi-\phi^\prime\bigr)
+{1\over2}\,\lambda^2\,.\eqno(5.1)$$

\bigskip\noindent
The canonical momenta are the same as in the other case, namely

$$\eqalignno{p&=0\,,&(5.2)\cr
\pi&=\dot\phi+\lambda\,.&(5.3)\cr}$$

\bigskip\noindent
The relation (5.2) is a (primary) constraint. Consistency conditions
lead to  secondary and tertiary constraints. These are

$$\eqalignno{&\pi-\phi^\prime=0&(5.4)\,,\cr
&\lambda^\prime=0\,.&(5.5)\cr}$$

\bigskip\noindent
It is important to mention that constraint (5.5) was just obtained by
virtue of the extra term $\lambda^2/2$ of the Lagrangian (5.1). We
rewrite all the constraints above as

$$\eqalignno{T_1&=p\,,&(5.6a)\cr
T_2&=\pi-\phi^\prime\,,&(5.6b)\cr
T_3&=\lambda^\prime\,.&(5.6c)\cr}$$

\bigskip\noindent
These are second-class and satisfy the algebra

$$\eqalignno{\bigl\{T_1(x),\,T_3(y)\bigr\}&=\delta^\prime(x-y)\,,\cr
\bigl\{T_2(x),\,T_2(y)\bigr\}&=-2\,\delta^\prime(x-y)\,.&(5.7)\cr}$$

\bigskip\noindent
It is opportune to mention that this theory, contrarily to what occurs
with the one discussed in Sec.~IV, has just one physical degree
of freedom (the same of the FJ theory).

\medskip
The direct use of the BFT procedure, as it was discussed in the
previous sections, leads to new (first-class) constraints and a
new canonical Hamiltonian. We just write them below

$$\eqalignno{\tilde T_1&=p+p_\chi\,,&(5.8a)\cr
\tilde T_2&=\pi-\phi^\prime+\psi\,,&(5.8b)\cr
\tilde T_3&=\lambda^\prime-\chi^\prime\,,&(5.8c)\cr
\tilde{\cal H}_{\rm C}&={1\over2}\,\bigl(\pi^2+\phi^{\prime2}\bigr)
+\bigl(\chi-\lambda\bigr)\bigl(\pi-\phi^\prime+\psi\bigr)
+{1\over2}\,\psi\,\bigl(\pi-\phi^\prime\bigr)
+{1\over4}\,\psi^2\,.&(5.9)\cr}$$

\bigskip\noindent
The auxiliary fields we have introduced are $\chi$ and its momentum
$p_\chi$ plus a constrained field $\psi$, which we have considered to
satisfy the same bracket structure given by (3.5) and (3.11). Using
the path integral formalism in the Faddeev-Senjanovic procedure, we
get

$$\eqalignno{Z=N\int[d\phi][d\psi]\delta[\tilde\chi]\,
\det\vert\{,\}\vert\exp\Bigl\{i\int d^2x\,
\Bigl[\phi^\prime\dot\phi&-\phi^{\prime2}
+\psi\bigl(\phi^\prime-\dot\phi\bigr)
-{1\over4}\psi^2\cr
&+{1\over4}\dot\psi\int dy\,\theta(x-y)\psi(y)
\Bigr]\Bigr\}\,.&(5.10)\cr}$$

\bigskip\noindent
We see that the effective Lagrangian extracted from (5.10) is exactly
the same as (3.27).

\eject
{\titlea VI. Conclusion}
\bigskip
We have studied chiral-boson theories by means of the BFT
quantization method. First we have considered the FJ formulation,
where we had to use just one constrained auxiliary variable in order
to convert the (continuous) second-class constraint into first-class.
As a result we have obtained a nonlocal Lagrangian, which is in
agreement with a previous treatment by means of Fourier modes
expansion.  Secondly we have dealt with the case with linear
constraint. We have obtained an effective Lagrangian that does not
lead to the FJ one when the auxiliary fields are turned off. This
fact may be reflecting the inconsistencies of the model as it was
initially formulated.  Finally, we have considered an improved way of
introducing the linear constraint in the chiral-boson theory. We have
shown that the inconsistencies have disappeared in this last
situation and that the BFT treatment leads to same result of the FJ
case.

\vskip 1cm
{\titlea Acknowledgment}
\bigskip
This work was supported in part by Conselho Nacional de
Desenvolvimento Cient\\fi- co e Tecnol\'ogico - CNPq (Brazilian
Research Agency).

\vfill\eject
{\titlea References}
\bigskip
\item {[1]} R. Floreanini and R. Jackiw, Phys. Rev. Lett. 59 (1987)
1873.
\item {[2]} P.A.M. Dirac, Can. J. Math. 2 (1950) 129; {\it Lectures
on quantum mechanics} (Yeshiva University, New York, 1964).
\item {[3]} I.R. Reyes-Martinez, J.R. Zimerman and T. Sobrinho, Phys.
Rev. D39 (1989) 3055.
\item {[4]} P.P. Srivastava, Phys. Rev. Lett. 63 (1989) 2791.
\item {[5]} K. Harada, Phys. Rev. Lett. 65 (1990) 267; H.O. Girotti,
M. Gomes and V.O. Rivelles, Phys. Rev. D45 (1992) R3329.
\item {[6]} J. Barcelos-Neto and C. Wotzasek, Europhys. Lett. 21
(1993) 511.
\item {[7]} W.T. Kim, J.K. Kim and Y.J. Park, Phys. Rev. D44 (1991)
563.
\item {[8]} W. Siegel, Nucl. Phys. B 238 (1984) 307.
\item {[9]} See for example M. Henneaux and C. Teitelboim, {\it
Quantization of gauge systems} (Princeton University Press, New
Jersey, 1992) and references therein.
\item {[10]} I.A. Batalin and E.S. Fradkin, Phys. Lett. B180 (1986)
157; Nucl. Phys. B279 (1987) 514;
\item {[11]} I.A. Batalin and I.V. Tyutin, Int.  J. Mod. Phuys. A6
(1991) 3255.
\item {[12]} R. Amorim and J. Barcelos-Neto, {\it BFT quantization of
the FJ chiral-boson}, to appear in Phys. Lett. B.
\item {[13]} Y.-G. Miao, J.-G. Zhou and Y.-Y. Liu, Phys. Lett. B323
(1994) 169.
\item {[14]} L.D. Faddeev, Theor. Math. Phys. 1 (1970) 1.
\item {[15]} P. Senjanovick, Ann. Phys. (N.Y.) 100 (1976) 277.
\item {[16]} C. Wotzasek, Phys. Rev. Lett. 66 (1991) 129.

\bye